\documentclass[sigconf]{acmart}
\usepackage{stfloats}

\AtBeginDocument{%
  \providecommand\BibTeX{{%
    \normalfont B\kern-0.5em{\scshape i\kern-0.25em b}\kern-0.8em\TeX}}}

\setcopyright{rightsretained}

\copyrightyear{2021}
\acmYear{2021}
\setcopyright{rightsretained}
\acmConference[WSDM '22]{Proceedings of the Fourteenth ACM International Conference on Web Search and Data Mining}{February 21--25, 2022}{Phoenix, Arizona}
\acmBooktitle{Proceedings of the Fourteenth ACM International Conference on Web Search and Data Mining (WSDM '22), February 21--25, 2022, Phoenix, Arizona}\acmDOI{10.1145/3437963.3441841}
\acmISBN{978-1-4503-8297-7/21/03}

\begin{document}

\title{Integrity 2022: Integrity in Social Networks and Media}

\author{Lluís Garcia-Pueyo}
\affiliation{%
  \institution{Facebook}
  \city{Menlo Park}
  \country{USA}
}
\email{lgp@fb.com}

\author{Panayiotis Tsaparas}
\affiliation{%
  \institution{University of Ioannina}
  \city{Ioannina}
  \country{Greece}
}
\email{tsap@cs.uoi.gr}

\author{Anand Bhaskar}
\affiliation{%
  \institution{Facebook}
  \city{Menlo Park}
  \country{USA}
}
\email{anandb@fb.com}

\author{Prathyusha Senthil Kumar}
\affiliation{%
  \institution{Facebook}
  \city{Menlo Park}
  \country{USA}
}
\email{prathyushas@fb.com}

\author{Roelof van Zwol}
\affiliation{%
  \institution{Pinterest}
  \city{San Francisco}
  \country{USA}
}
\email{roelof@pinterest.com}

\author{Timos Sellis}
\affiliation{%
  \institution{Facebook}
  \city{Menlo Park}
  \country{USA}
}
\email{tsellis@fb.com}

\author{Anthony McCosker}
\affiliation{%
  \institution{Swinburne Social Innovation Research Institute}
  \city{Melbourne}
  \country{Australia}
}
\email{amccosker@swin.edu.au}

\author{Paolo Papotti}
\affiliation{%
  \institution{EURECOM}
  \city{Biot}
  \country{France}
}
\email{paolo.papotti@eurecom.fr}

\renewcommand{\shortauthors}{Garcia-Pueyo, et al.}

\begin{abstract}
This is the proposal for the third edition of the Workshop on Integrity in Social Networks and Media, Integrity 2022, following the success of the first two Workshops held in conjunction with the 13th \& 14th ACM Conference on Web Search and Data Mining (WSDM) in 2020~\cite{Integrity2020} and 2021~\cite{Integrity2021}. The goal of the workshop is to bring together researchers and practitioners to discuss content and interaction integrity challenges in social networks and social media platforms. The event consists of (1) a series of invited talks by reputed members of the Integrity community from both academia and industry, (2) a call-for-papers for contributed talks and posters, and (3) a panel with the speakers.
\end{abstract}


\maketitle

\newpage
\section{Workshop Description}
Integrity 2022 aims to repeat the success achieved in the previous two workshop editions, Integrity 2020 \& 2021, hosted in the corresponding years of the WSDM conference. Previous workshops have featured invited talks from industry leads from companies like Facebook, Twitter, Pinterest, LinkedIn, Snap, and Airbnb, along with academic experts from institutions like UC Berkeley, MIT, Princeton, Aalto, and Queensland University of Technology. These talks have exposed challenges, solutions, and ongoing research in areas such as Misinformation, Bias in Machine Learning models, content- and behavior-based detection of quality problems, Display Advertising Integrity, Safety vs Privacy, Opinion Dynamics, and several other topics.

The previous workshops have resulted in fruitful discussions and engagement from the audience, and a unanimous push towards organizing a recurring workshop exploring these problems and potential solutions, with participation from academics and industry researchers. Besides, there is a strong interest in the community in integrity, with several related workshops and conferences on related topics~\cite{Fairness, Rome, Cibersafety}.

In the past decade, social networks and social media sites, such as Facebook and Twitter, have become the default channels of communication and information. The popularity of these online portals has exposed a collection of integrity issues: cases where the content produced and exchanged compromises the quality, operation, and eventually the integrity of the platform. Examples include misinformation, low quality and abusive content and behaviors, and polarization and opinion extremism. There is an urgent need to detect and mitigate the effects of these integrity issues, in a timely, efficient, and unbiased manner.

This workshop aims to bring together top researchers and practitioners from academia and industry, to engage in a discussion about algorithmic and systems aspects of integrity challenges. The WSDM Conference, that combines Data Mining and Machine Learning with research on Web and Information Retrieval offers the ideal forum for such a discussion, and we expect the workshop to be of interest to everyone in the community. The topic of the workshop is also interdisciplinary, as it overlaps with psychology, sociology, and economics, while also raising legal and ethical questions, so we expect it to attract a broad audience.

As indicated by the organizing committee and the speaker list, the workshop aims to bring together researchers and practitioners from both industry and academia, leading to exchange of knowledge and cross-cutting collaborations. The event consists of a series of invited talks by reputed members of the Integrity community from both academia and industry, contributed talks or posters, and a panel with the speakers.

The workshop topics include, but are not limited to:
\begin{itemize}
\item \textbf{Low quality, borderline, and offensive content and behaviors:} Methods for detecting and mitigating low quality and offensive content and behaviors, such as clickbait, fake engagement, nudity and violence, bullying, and hate speech.
\item \textbf{Personalized treatment of low quality content:} Identification, measurement, and reduction of bad experiences.
\item \textbf{COVID-19 on social media:} Authoritative health information; Covid misinformation; Vaccine hesitancy; Anti-vax movements.
\item \textbf{Misinformation:} Detecting and combating misinformation; Prevalence and virality of misinformation; Misinformation sources and origins; Source and content credibility; Inoculation strategies; Deep and shallow fakes.
\item \textbf{Polarization:} Models and metrics for polarization; Echo chambers and filter bubbles; Opinion Extremism and radicalization; Algorithms for mitigating polarization.
\item \textbf{Fairness in Integrity:} Fairness in the detection and mitigation of integrity issues with respect to sensitive attributes such as gender, race, sexual orientation, and political affiliation.
\end{itemize}

\begin{table*}[t]
  \caption{Workshop schedule}
  \label{tab:schedule}
  \begin{tabular}{cl}
    \toprule
    Time &Event \\
    \midrule
    8:45 & Welcoming remarks \\
    9:00 & Invited Talk - Academic \\
    9:30 & Invited Talk - Industrial \\
    10:00 & Coffee Break \\
    10:30 & Contributed Talks/Posters \\
    12:00 & Lunch \\
    13:00 & Invited Talk - Academic \\
    13:30 & Invited Talk - Industrial \\
    14:00 & Coffee Break \\
    14:30 & Contributed Talks/Posters \\
    16:00 & Panel with Speakers \& Closing remarks \\
    \bottomrule
  \end{tabular}
\end{table*}

\section{Workshop Format}
\begin{itemize}
\item \textbf{Duration \& Format:} Full day, invited and contributed speakers.
\item \textbf{Number of participants (estimated):} 40
\item \textbf{Schedule (tentative):} Table~\ref{tab:schedule} details the proposed schedule.
\item \textbf{Call-for-papers, Poster session and Contributed talks:} The committee will open the workshop to additional contributors through contributed talks and a poster session, via a call-for-papers. The schedule will be adapted based on the contributed material.
\end{itemize}

\section{Organizers}
\textbf{Lluis Garcia-Pueyo}, Facebook, is an Engineering Manager at Facebook, leads the News Feed Integrity Distribution pillar focusing on personalization, discovery and reduction of negative experiences in News Feed and Stories ranking. Prior to this, he worked in information extraction and information retrieval at Google Research, and multimedia retrieval and display advertising at Yahoo Research. Lluis holds an MS in Computer Science from the Universitat Politècnica de Catalunya. His research has been published in top-tier conferences such as WWW, KDD, SIGIR, ACM Multimedia, and WSDM, and he is a member of the PC for KDD, WWW and other conferences. He is an organizer of the internal Integrity Week conference at Facebook, which hosts 200+ research scientists, data scientists and engineers in Social Network Integrity topics. He organized the Integrity 2020 and 2021 editions of this workshop at WSDM’20 \& '21. \\

\textbf{Panayiotis Tsaparas}, University of Ioannina, received his Ph.D. from the University of Toronto in 2003. Since then he has worked as a postdoctoral fellow at the University of Rome, “La Sapienza”, and at the University of Helsinki, and as a researcher at Microsoft Research. Since 2011 he joined the Department of Computer Science and Engineering of University of Ioannina, where he is now an Associate Professor.  His research interests include Social Network Analysis, Data Mining, Machine Learning, and Information Retrieval. He has served several times as a PC and Senior PC member for premier Data Mining and Data Bases conferences such as KDD, WWW, WSDM, VLDB, ICDE, and as a reviewer for journals such as TKDE, TWEB, CACM, TODS, KAIS, DMKD, while he is an associate editor for the TKDE and OSNM Journal. He was in the organizing committee of the OCeANS workshop at KDD 2018. He has served in 3 NSF panels, and as a reviewer for Hellenic Research Foundation. During his tenure at Microsoft he received 3 technology transfer awards for successful transfer of research results to product groups. He has published 60 papers in peer-reviewed conferences and journals, and has filed for 12 patents, 8 of which have been awarded. He organized the Integrity 2020 workshop collocated within WSDM’20~\cite{Integrity2020}. \\

\textbf{Anand Bhaskar}, Facebook. Anand Bhaskar is a Research Scientist at Facebook, where he works on building and incorporating content quality signals into News Feed ranking and studying the network effects of ranking changes. Prior to that, he was a postdoctoral researcher at Stanford University and HHMI, where he applied techniques from statistics, computer science, and applied mathematics to large-scale genomic datasets for addressing scientific questions such as the genetic basis of disease, human demographic history, and forensics, among others. His work has been published in journals such as PNAS and the Annals of Statistics, and conferences such as VLDB and SIGMOD. He received a Ph.D. in Computer Science and an M.A. in Statistics from the University of California, Berkeley, and M.Eng. and B.S. degrees in Computer Science from Cornell University. His research has been supported by a Berkeley Fellowship, Simons-Berkeley Research Fellowship, Japan Society for the Promotion of Science Postdoctoral Fellowship, and Stanford CEHG Postdoctoral Fellowship. He organized the Integrity 2020 and 2021 editions of this workshop at WSDM’20 \& '21. \\

\textbf{Prathyusha Senthil Kumar}, Facebook, is an Engineering Manager at Facebook, where she leads News Feed Integrity efforts leveraging machine learning techniques to understand and utilize content quality in feed ranking and to reduce subjective bad experiences through personalized ranking interventions. Prior to joining Facebook, she led the Search Query Understanding and Relevance Ranking applied research teams at Ebay Inc. Prior to that she worked as a researcher at EBay applying machine learning, natural language processing and information retrieval for various search problems. She holds a master’s degree in Computer Science from the University of Texas at Austin and an undergraduate degree in Information Technology from the College of Engineering Guindy, Anna University. Her research has been published in conferences such as SIGIR, CIKM and IEEE. \\

\textbf{Roelof van Zwol} is the head of Ads Quality at Pinterest. The team is responsible for (1) helping advertisers define the audience they want to reach through services such as Act-alike modeling, interest targeting, etc, as well as the ML models that power the ads delivery system to determine which ads to show to a Pinner in a generalized second price auction. Previously, Roelof was the Director of Product Innovation at Netflix. There he was responsible for the innovation of Netflix’s content promotion and acquisition algorithms. Prior to joining Netflix, Roelof managed the multimedia research team at Yahoo!, first from Barcelona, Spain, and later from Yahoo!’s headquarters in California. He started his career in academia as an assistant professor in the Computer Science department in Utrecht, the Netherlands, after finishing his PhD at the University of Twente in Enschede, the Netherlands. He organized the Integrity 2021 workshop collocated within WSDM’21~\cite{Integrity2020}. \\

\textbf{Timos Sellis} is a visiting scientist at Facebook and an Adjunct Professor in Computer Science at Swinburne University of Technology, where he also served as the director of the Data Science Research Institute (2026-20). He received his M.Sc. degree from Harvard University (1983) and Ph.D. degree from the University of California at Berkeley (1986). He has served as a professor at the University of Maryland (1986-92), the National Technical University of Athens (1992-2013), and was the inaugural Director of the Institute for the Management of Information Systems of the "Athena" Research Center (2007-13).He is IEEE Fellow (2009) and an ACM Fellow (2013), for his contributions to database systems and data management. In March 2018 he received the IEEE TCDE Impact Award, for contributions to database systems research and broadening the reach of data engineering research. \\

\textbf{Anthony McCosker} is Associate Professor of media and communication, Deputy Director of the Swinburne University’s Social Innovation Research institute, and Chief Investigator in the ARC Centre of Excellence for Automated Decision Making and society. He researches the impact and uses of social media, data and new communication technologies, with a focus on mental health, digital citizenship, inclusion and literacy. He is co-author of the book Automating Vision: The Social Impact of the New Camera Consciousness (Routledge), and co-author of the forthcoming book Everyday Data Cultures (Polity Press). \\

\textbf{Paolo Papotti} got his Ph.D. degree from the University of Roma Tre (Italy) in 2007 and is an associate professor in the Data Science department at EURECOM (France) since 2017. Before joining EURECOM, he has been a scientist in the data analytics group at QCRI (Qatar) and an assistant professor at Arizona State University (USA). His research is in the broad areas of scalable data management and information quality, with a focus on data integration and computational claim verification.

\section{Related Workshops}
\subsection{Integrity 2020 \& 2021}
Hosted in WSDM 2020 at Houston, TX\footnote{https://sites.google.com/view/integrity-workshop} and WSDM 2021 in an online event\footnote{http://integrity-workshop.org/}, the previous editions of this workshop~\cite{Integrity2020, Integrity2021} brought together Integrity experts from industry leaders with researchers, and focussed on content-based integrity, integrity and abuse in display advertising, misinformation, behavioral analysis, and integrity challenges for machine learning applications.

\subsection{Misinfo 2021}
The Workshop on Misinformation Integrity in Social Networks\footnote{https://sites.google.com/view/misinfo-2021}\cite{Misinfo2021}, held in conjunction with The Web Conference 2021, focused on topics related to detecting, measuring, and mitigating misinformation and polarization.

\subsection{CyberSafety 2019}
Hosted In The Web Conference 2019\footnote{https://cybersafety2019.github.io/}, this workshop focussed on anomalous behaviors such as fraudulent engagement, misinformation and propaganda, user deception and scams, harassment, hate speech, cyberthreats, cyberbullying on social networks.

\subsection{MisinfoWorkshop2019}
The International Workshop on Misinformation, Computational Fact-Checking and Credible Web, in The Web Conference 2019\footnote{https://sites.google.com/view/misinfoworkshop}, focused on computational methodology for Misinformation and fact-checking detection, and Ethical pitfalls and solutions, as well as Education on Misinformation.

\subsection{FATES on the Web’19}
Hosted by The Web Conference 2019, this workshop, with a focus on Social Sciences, discussed issues such as Transparency, Credibility, Fairness, Bias and Ethics in computational research and analysis.

\subsection{OCeANS Workshop’18}
The Opinions, Conflict, and Abuse in a Networked Society\footnote{https://sites.google.com/view/oceans-kdd2018/home}, hosted in ACM SIGKDD’18, had talks on crowdsourcing and the effects of the usage of user data in detection tasks, methods for low-quality content detection, and adversarial system design.

\subsection{ROME 2019}
The Workshop on Reducing Online Misinformation Exposure\footnote{https://rome2019.github.io/}, colocated with SIGIR 2019, presented work on subjectivity on crowdsourcing, credibility and bias, medical misinformation, user-generated video verification and time-sensitive fact-checking.

\subsection{Truth Discovery 2019}
The Truth Discovery and Fact Checking: Theory and Practice\footnote{https://truth-discovery-kdd2019.github.io/} workshop, collocated with ACM SIGKDD’19, discussed a broad range of topics related to the Misinformation discovery problem: general architectures, natural language processing for claims, using fact-checking in supervised settings, information extraction, plagiarism.

\bibliographystyle{ACM-Reference-Format}
\bibliography{sample-base}


\begin{thebibliography}{6}


\ifx \showCODEN    \undefined \def \showCODEN     #1{\unskip}     \fi
\ifx \showDOI      \undefined \def \showDOI       #1{#1}\fi
\ifx \showISBNx    \undefined \def \showISBNx     #1{\unskip}     \fi
\ifx \showISBNxiii \undefined \def \showISBNxiii  #1{\unskip}     \fi
\ifx \showISSN     \undefined \def \showISSN      #1{\unskip}     \fi
\ifx \showLCCN     \undefined \def \showLCCN      #1{\unskip}     \fi
\ifx \shownote     \undefined \def \shownote      #1{#1}          \fi
\ifx \showarticletitle \undefined \def \showarticletitle #1{#1}   \fi
\ifx \showURL      \undefined \def \showURL       {\relax}        \fi
\providecommand\bibfield[2]{#2}
\providecommand\bibinfo[2]{#2}
\providecommand\natexlab[1]{#1}
\providecommand\showeprint[2][]{arXiv:#2}

\bibitem[\protect\citeauthoryear{Albarghouthi and Vinitsky}{Albarghouthi and
  Vinitsky}{2019}]%
        {Fairness}
\bibfield{author}{\bibinfo{person}{Aws Albarghouthi} {and}
  \bibinfo{person}{Samuel Vinitsky}.} \bibinfo{year}{2019}\natexlab{}.
\newblock \showarticletitle{Fairness-Aware Programming}. In
  \bibinfo{booktitle}{\emph{Proceedings of the Conference on Fairness,
  Accountability, and Transparency}} (Atlanta, GA, USA)
  \emph{(\bibinfo{series}{FAT* '19})}. \bibinfo{publisher}{Association for
  Computing Machinery}, \bibinfo{address}{New York, NY, USA},
  \bibinfo{pages}{211–219}.
\newblock
\showISBNx{9781450361255}
\urldef\tempurl%
\url{https://doi.org/10.1145/3287560.3287588}
\showDOI{\tempurl}


\bibitem[\protect\citeauthoryear{Bouchard, Caldarelli, and Plachouras}{Bouchard
  et~al\mbox{.}}{2019}]%
        {Rome}
\bibfield{author}{\bibinfo{person}{Guillaume Bouchard}, \bibinfo{person}{Guido
  Caldarelli}, {and} \bibinfo{person}{Vassilis Plachouras}.}
  \bibinfo{year}{2019}\natexlab{}.
\newblock \showarticletitle{ROME 2019: Workshop on Reducing Online
  Misinformation Exposure}. In \bibinfo{booktitle}{\emph{Proceedings of the
  42nd International ACM SIGIR Conference on Research and Development in
  Information Retrieval}} (Paris, France) \emph{(\bibinfo{series}{SIGIR'19})}.
  \bibinfo{publisher}{Association for Computing Machinery},
  \bibinfo{address}{New York, NY, USA}, \bibinfo{pages}{1426–1428}.
\newblock
\showISBNx{9781450361729}
\urldef\tempurl%
\url{https://doi.org/10.1145/3331184.3331645}
\showDOI{\tempurl}


\bibitem[\protect\citeauthoryear{Garcia-Pueyo, Bhaskar, Kumar, Tsaparas,
  Garimella, Sun, and Bonchi}{Garcia-Pueyo et~al\mbox{.}}{2021a}]%
        {Misinfo2021}
\bibfield{author}{\bibinfo{person}{Llu\'{\i}s Garcia-Pueyo},
  \bibinfo{person}{Anand Bhaskar}, \bibinfo{person}{Prathyusha~Senthil Kumar},
  \bibinfo{person}{Panayiotis Tsaparas}, \bibinfo{person}{Kiran Garimella},
  \bibinfo{person}{Yu Sun}, {and} \bibinfo{person}{Francesco Bonchi}.}
  \bibinfo{year}{2021}\natexlab{a}.
\newblock \showarticletitle{MISINFO 2021: Workshop on Misinformation Integrity
  in Social Networks}.
\newblock  (\bibinfo{year}{2021}).
\newblock


\bibitem[\protect\citeauthoryear{Garcia-Pueyo, Bhaskar, Tsaparas, Gionis,
  Eliassi-Rad, Daltayanni, Sun, and Papadimitriou}{Garcia-Pueyo
  et~al\mbox{.}}{2020}]%
        {Integrity2020}
\bibfield{author}{\bibinfo{person}{Llu\'{\i}s Garcia-Pueyo},
  \bibinfo{person}{Anand Bhaskar}, \bibinfo{person}{Panayiotis Tsaparas},
  \bibinfo{person}{Aristides Gionis}, \bibinfo{person}{Tina Eliassi-Rad},
  \bibinfo{person}{Maria Daltayanni}, \bibinfo{person}{Yu Sun}, {and}
  \bibinfo{person}{Panagiotis Papadimitriou}.} \bibinfo{year}{2020}\natexlab{}.
\newblock \showarticletitle{Integrity 2020: Integrity in Social Networks and
  Media}. In \bibinfo{booktitle}{\emph{Proceedings of the 13th International
  Conference on Web Search and Data Mining}} (Houston, TX, USA)
  \emph{(\bibinfo{series}{WSDM '20})}. \bibinfo{publisher}{Association for
  Computing Machinery}, \bibinfo{address}{New York, NY, USA},
  \bibinfo{pages}{905–906}.
\newblock
\showISBNx{9781450368223}
\urldef\tempurl%
\url{https://doi.org/10.1145/3336191.3371880}
\showDOI{\tempurl}


\bibitem[\protect\citeauthoryear{Garcia-Pueyo, Bhaskar, van Zwol, Sellis,
  Ranade, Senthil~Kumar, Sun, and Zhang}{Garcia-Pueyo et~al\mbox{.}}{2021b}]%
        {Integrity2021}
\bibfield{author}{\bibinfo{person}{Llu\'{\i}s Garcia-Pueyo},
  \bibinfo{person}{Anand Bhaskar}, \bibinfo{person}{Roelof van Zwol},
  \bibinfo{person}{Timos Sellis}, \bibinfo{person}{Gireeja Ranade},
  \bibinfo{person}{Prathyusha Senthil~Kumar}, \bibinfo{person}{Yu Sun}, {and}
  \bibinfo{person}{Joy Zhang}.} \bibinfo{year}{2021}\natexlab{b}.
\newblock \showarticletitle{Integrity 2021: Integrity in Social Networks and
  Media}. In \bibinfo{booktitle}{\emph{Proceedings of the 14th ACM
  International Conference on Web Search and Data Mining}} (Virtual Event,
  Israel) \emph{(\bibinfo{series}{WSDM '21})}. \bibinfo{publisher}{Association
  for Computing Machinery}, \bibinfo{address}{New York, NY, USA},
  \bibinfo{pages}{1159–1160}.
\newblock
\showISBNx{9781450382977}
\urldef\tempurl%
\url{https://doi.org/10.1145/3437963.3441841}
\showDOI{\tempurl}


\bibitem[\protect\citeauthoryear{Han and Shah}{Han and Shah}{2019}]%
        {Cibersafety}
\bibfield{author}{\bibinfo{person}{Richard Han} {and} \bibinfo{person}{Neil
  Shah}.} \bibinfo{year}{2019}\natexlab{}.
\newblock \showarticletitle{Cybersafety 2019: The 4th Workshop on Computational
  Methods in Online Misbehavior}. In \bibinfo{booktitle}{\emph{Companion
  Proceedings of The 2019 World Wide Web Conference}} (San Francisco, USA)
  \emph{(\bibinfo{series}{WWW '19})}. \bibinfo{publisher}{Association for
  Computing Machinery}, \bibinfo{address}{New York, NY, USA},
  \bibinfo{pages}{146–147}.
\newblock
\showISBNx{9781450366755}
\urldef\tempurl%
\url{https://doi.org/10.1145/3308560.3316493}
\showDOI{\tempurl}


\end{thebibliography}

\end{document}